%% file: mcglatencypaper.tex
\documentclass{sig-alternate-05-2015}

\usepackage{tikz}
\usepackage{bm,array}

\graphicspath{{./figures/}}

\begin{document}

\newcolumntype{C}{>{\centering\arraybackslash}p{2em}}

\setcopyright{acmcopyright}

\doi{10.475/123_4}

\isbn{123-4567-24-567/08/06}

\conferenceinfo{HotMobile '16}{February 21--22, 2017, Sonoma, California, USA}

\acmPrice{\$15.00}

%

\title{Dissecting the End-to-end Latency of Interactive Mobile Video Applications}
%
%
%
%
%

\numberofauthors{2} 
%
\author{
%
%
\alignauthor
Teemu K\"am\"ar\"ainen, Matti Siekkinen, Antti Yl\"a-J\"a\"aski\\
       \affaddr{Department of Computer Science}\\
       \affaddr{Aalto University, Finland}\\
       \email{firstname.lastname@aalto.fi}
\alignauthor Wenxiao Zhang, Pan Hui\\
       \affaddr{Systems and Media Lab}\\
       \affaddr{The Hong Kong University of Science and Technology, Hong Kong}\\
       \email{\mbox{wzhangal@stu.ust.hk,~panhui@cse.ust.hk}}
\and  
}

\maketitle
\begin{abstract}
\input{mcglatencypaper-abstract}
\end{abstract}

%
%


%
%

%
%



\section{Introduction}
\input{mcglatencypaper-introduction}
\section{Measurement setup}
\input{mcglatencypaper-measurementsetup}
\section{RGR: Mobile Cloud Gaming}
\input{mcglatencypaper-delayanalysis-cg}
\section{Mobile Virtual Reality}
\input{mcglatencypaper-delayanalysis-vr}
\section{Mobile Augmented Reality}
\input{mcglatencypaper-delayanalysis-ar}
\section{Discussion of results}
\input{mcglatencypaper-discussion}
\section{Conclusion}
\input{mcglatencypaper-conclusion}


%
\bibliographystyle{abbrv}
\bibliography{IEEEabrv,mcglatencypaper} 
%
%
\end{document}

%% file: mcglatencypaper-abstract.tex
In this paper we measure the step-wise latency in the pipeline of three kinds of interactive mobile video applications that are rapidly gaining popularity, namely Remote Graphics Rendering (RGR) of which we focus on mobile cloud gaming, Mobile Augmented Reality (MAR), and Mobile Virtual Reality (MVR). The applications differ from each other by the way in which the user interacts with the application, i.e., video I/O and user controls, but they all share in common the fact that their user experience is highly sensitive to end-to-end latency. Long latency between a user control event and display update renders the application unusable. Hence, understanding the nature and origins of latency of these applications is of paramount importance. We show through extensive measurements that control input and display buffering have a substantial effect on the overall delay. Our results shed light on the latency bottlenecks and the maturity of technology for seamless user experience with these applications.

%% file: mcglatencypaper-introduction.tex
In recent years, new ways to bring interactive multimedia to mobile devices have emerged. Examples include Remote Graphics Rendering (RGR), Mobile Augmented Reality (MAR), and Mobile Virtual Reality (MVR) applications. They all commonly leverage the superior capability of cloud computing to render graphics and process video but their I/O and user interactions differ. One of the biggest challenges with these applications is that their user experience may degrade dramatically if it takes even just a few hundred milliseconds to update the display after a control action by the user. Hence, understanding and optimizing the end-to-end latency in user interactions with these applications is critically important.

In what we call RGR applications, the user runs on a thin client software that intercepts user control events and sends them to the cloud. The application logic is executed and graphics rendered completely on the cloud and the client typically receives back a video stream. Cloud gaming and full-cloud CAD are examples of RGR applications. In MAR applications, the input stream is usually a camera feed from the mobile device. This video is either streamed to a cloud or processed locally, depending on the application and device capabilities. Video processing recognizes and tracks features of interest and additional (augmented) objects or information are drawn to the screen of the mobile device. MVR applications are usually used together with a headset to render a complete virtual world for the user. Head tracking using the sensors of the mobile phone and possibly a remote controller are the input to the application and the mobile phone renders the resulting projection of the virtual world depending on the user's head movements.

The added delay of rendering, processing, encoding, decoding, and transmitting video through the network are key factors that affects the user experience. In addition, the different user interactions, i.e., touch screen, separate gamepad, or sensors, must be accounted for. To minimize the end-to-end latency, we must first precisely understand where in the pipeline it accumulates, and that is the objective of our work. Previous studies have attempted to quantify some of the latency components of such applications, but their methodology has been limited and most of them have not considered scenarios involving mobile devices. They have mostly used either timing hooks injected directly into the code or a high-speed camera\cite{chen2014TMM,ivkovic2015chi,chang2016mm}. Timing hooks are useful for measuring independent tasks occurring completely in the mobile device, whereas a high-speed camera can only capture the total delay from a control press to display update. A predictive approach has also been proposed by Cattan et al. but it requires separate calibration\cite{cattan2015its}. A more detailed and precise break down of delay components requires new methods.

In this paper we utilize a modified and extended version of the WALT Latency Timer \cite{google_walt} together with code injections dissect the latency of these mobile applications into subcomponents. Similar measurement setups have been previously utilized in measuring mobile phone display responsiveness \cite{beyer2015method,deber2016hammer}. Our approach can also measure gyro, gamepad and Bluetooth delay on top of the traditional touch latency measurements. Our setup is also in-sync with the mobile phone, allowing a more precise division of latency components. Our results reveal where in the processing pipeline of these applications lie the major latency bottlenecks and how different control methods affect the end-to-end latency. Based on the results, we also discuss whether the technology is mature enough for seamless user experience with these applications and highlight the most promising avenues for latency.

%% file: mcglatencypaper-measurementsetup.tex
\begin{figure}
\centering
\begin{tikzpicture}
    \node[anchor=south west,inner sep=0] (image) at (0,0,0) {\includegraphics[width=0.75\columnwidth]{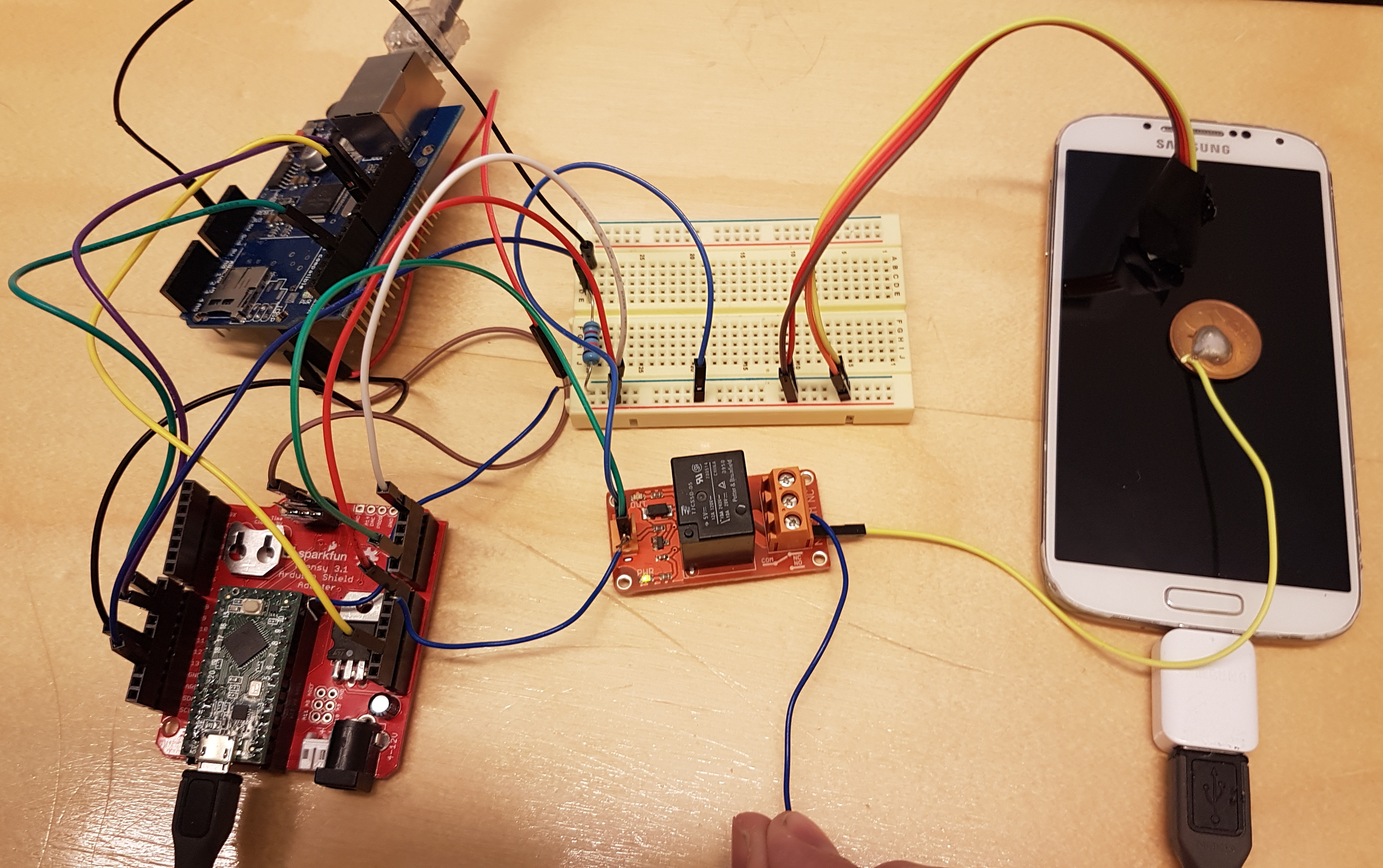}};
    \begin{scope}[x={(image.south east)},y={(image.north west)}]
        \draw[dashed,-latex] (0.14,0.14) -- +(-0.53in,0)node[anchor=east] {1}; 
        \draw[dashed,-latex] (0.875,0.625) -- +(0.48in,0)node[anchor=west] {2}; 
        \draw[dashed,-latex] (0.5,0.4) -- +(1.41in,0)node[anchor=west] {3}; 
        \draw[dashed,-latex] (0.55,0.05) -- +(-1.56in,0)node[anchor=east] {4}; 
	\draw[dashed,-latex] (0.85,0.75) -- +(0.54in,0)node[anchor=west] {5}; 
        \draw[dashed,-latex] (0.2,0.75) -- +(-0.69in,0)node[anchor=east] {6}; 
    \end{scope}
\end{tikzpicture}
\caption{Measurement setup for touch, gamepad, Ethernet and screen delay analysis.} \label{fig:setup1}
\end{figure}
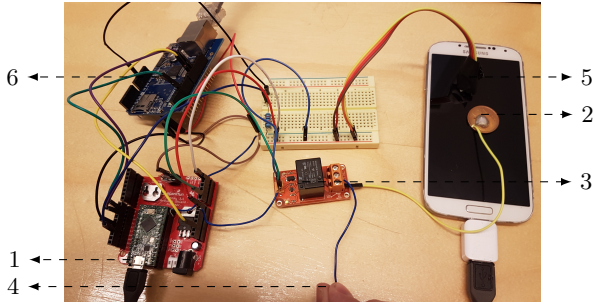

\label{chapter:measurementsetup}
The measurement setup depicted in Figure \ref{fig:setup1} uses an Arduino-compatible board (Teensy LC) connected to the phone through a USB connection (1). The Teensy board is configured to act as a joystick in addition to a serial connection through the USB. This enables us to programmatically enter key presses to the mobile device. To simulate touch presses, we use a coin (2) attached to a relay (3) which closes a connection loop to the human tester (4) when activated. This in turn enables us to precisely measure the time when a touch is initiated on the display.

In addition two BPW34 photodiodes (5) catch the time when a frame has been updated on the display of the mobile device. The photodiode can sense the change from a dark frame to a more illuminated frame, for example from color black to white. We utilize this property in the software to measure the display times of specific frames. The measurement board has also an Ethernet shield (6) attached for Internet connectivity. This is required to measure how long does it take to prepare and send a packet including a control event from the mobile device towards a server. This time period can be calculated by directing the mobile device to send the control command directly to the Teensy measurement board which also initiated the control command.

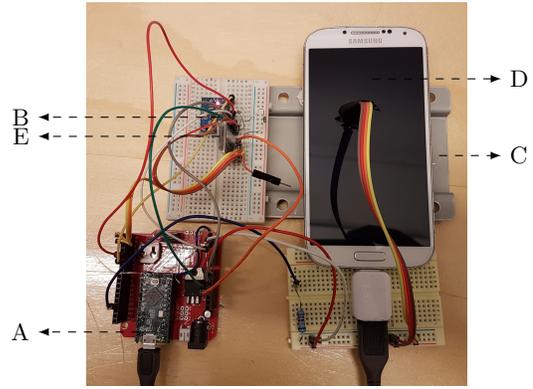
\begin{figure}
\centering
\begin{tikzpicture}
    \node[anchor=south west,inner sep=0] (image) at (0,0,0) {\includegraphics[width=0.6\columnwidth]{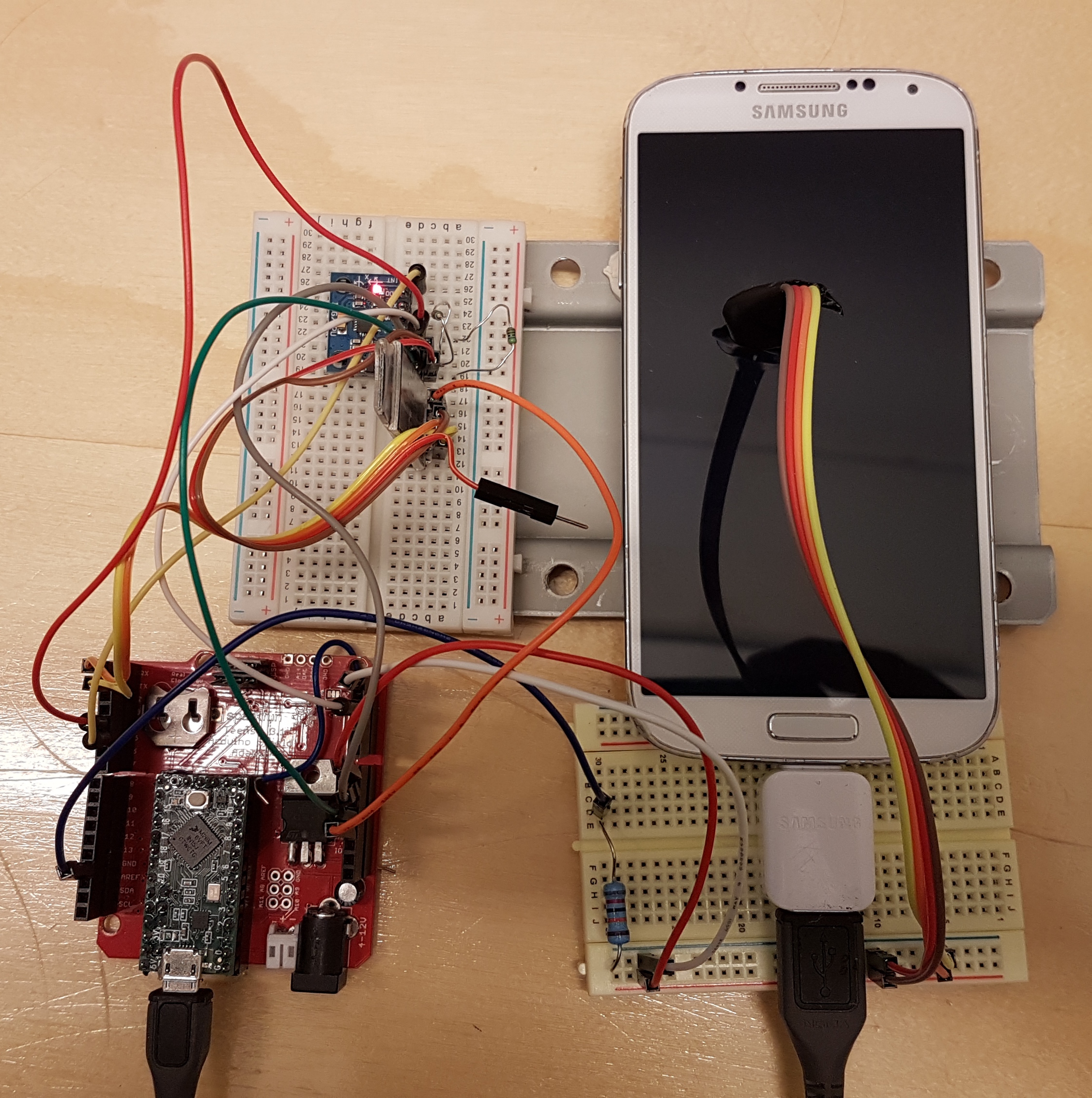}};
    \begin{scope}[x={(image.south east)},y={(image.north west)}]
        \draw[dashed,-latex] (0.14,0.14) -- +(-0.53in,0)node[anchor=east] {A}; 
	\draw[dashed,-latex] (0.35,0.7) -- +(-0.95in,0)node[anchor=east] {B}; 
        \draw[dashed,-latex] (0.925,0.6) -- +(0.325in,0)node[anchor=west] {C}; 
        \draw[dashed,-latex] (0.75,0.8) -- +(0.675in,0)node[anchor=west] {D}; 
        \draw[dashed,-latex] (0.35,0.65) -- +(-0.95in,0)node[anchor=east] {E}; 
    \end{scope}
\end{tikzpicture}
\caption{Measurement setup for gyro and Bluetooth delay analysis.} \label{fig:setup2}
\end{figure}

For the virtual reality application experiments a reference gyro value is needed to compare the responsiveness of the gyro sensor inside the mobile phone. The modified measurement setup is presented in Figure \ref{fig:setup2} Using the same Teensy-board (A), we attached also a MPU-6050 gyro sensor (B) to the measurement setup. We recorded raw gyro values during the experiment to get indication of movement as quickly as possible for a base reference value. The gyroscope sensor was attached to a plate (C) together with the mobile phone (D). When we moved the plate, both mobile phone and the gyro sensor moved simultaneously. The setup also includes a Bluetooth Low Energy module (E) for measuring the Bluetooth controller delay in an MVR application.

Without time synchronization, the Teensy board can calculate events that started and ended in the device itself. In addition the mobile device can calculate the delay of events occurring within the mobile device. However to calculate the delay of a control command which initiated in the Teensy side and was received in the mobile device, requires us to sync the times of the devices. This is possible with the WALT Latency timer which utilizes the fast USB connection of the Teensy board and an algorithm similar to NTP to sync the clocks of the devices.

We used three mobile devices in our test experiments: Samsung S4, Samsung S7 and Huawei Nexus 6P. Samsung S4 was released in 2013, Nexus 6P in 2015 and Samsung S7 in 2016. All of the devices are top of the line in performance when released so comparing the devices should give a hint on how the different delay components are developing and will develop in the future.

%% file: mcglatencypaper-delayanalysis-cg.tex
\begin{figure}[!t]
\centering
\includegraphics[width=\columnwidth]{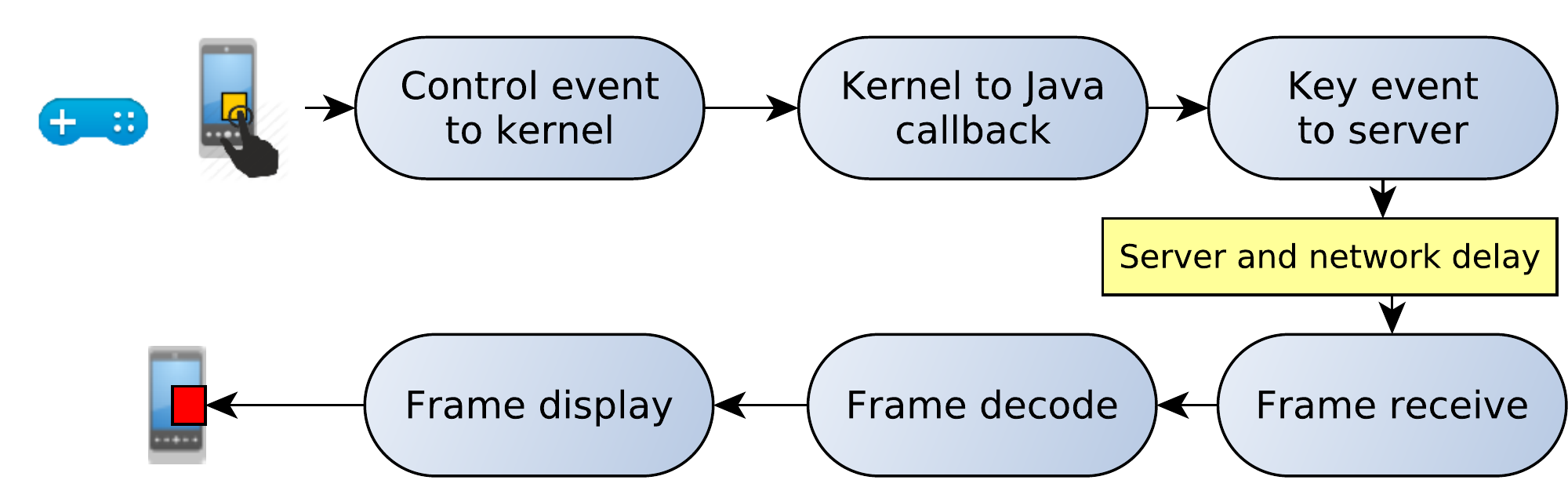}
\caption{Typical pipeline of a mobile cloud gaming scenario.}
\label{fig:cg_pipeline}
\end{figure}

Mobile cloud gaming is a prime example of an RGR application. It has very stringent requirements since it requires both low latency and high bandwidth for good quality of experience. 
We focus on the delays occurring on the client side as network and server delay have been covered in depth in existing literature. The complete pipeline of a typical cloud gaming scenario is presented in Figure \ref{fig:cg_pipeline}. The measurement results for the cloud gaming use case are presented in Table \ref{table:cgall}. We use the GamingAnywhere open-source cloud gaming application in our measurements.

\subsection{Control delay}
Control delay is the delay between the user initiating a command using the touch screen or by pressing a button on an external control device and the operating system of the mobile device registering the command. The control device can also be integrated as part of the device as is the case with for example Nvidia's Shield device. In this paper we measure both cases as the measurement setup can simulate both user touch interactions and gamepad commands.

\begin{table}[]
\centering
\caption{Cloud gaming delay measurement results.}
\label{table:cgall}
\begin{scriptsize}
\begin{tabular}{l|c|c|c|c|}
\cline{2-5}
                                                       & \multicolumn{2}{l|}{\textbf{Samsung S4}} & \multicolumn{2}{l|}{\textbf{Samsung S7}} \\ \cline{2-5} 
                                                       & \textbf{Avg.}      & \textbf{SD}     & \textbf{Avg.}      & \textbf{SD}     \\ \hline
\multicolumn{1}{|l|}{\textbf{Touch to kernel (ms)}}    & 40.5               & 2.3                 & 24.1               & 3.0                 \\ \hline
\multicolumn{1}{|l|}{\textbf{Gamepad to kernel (ms)}}  & 0.6                & 0.6                 & 0.2                & 0.4                 \\ \hline
\multicolumn{1}{|l|}{\textbf{Kernel to callback (ms)}} & 5.5                & 1.6                 & 3.4                & 0.6                 \\ \hline
\multicolumn{1}{|l|}{\textbf{Callback to radio (ms)}}  & 9.1                & 2.7                 & 1.6                & 0.8                 \\ \hline
\multicolumn{1}{|l|}{\textbf{Frame receive (ms)}}      & 10.5               & 5.8                 & 9.6                & 4.5                 \\ \hline
\multicolumn{1}{|l|}{\textbf{Frame decode (ms)}}       & 20.4               & 11.6                & 8.3                & 1.1                 \\ \hline
\multicolumn{1}{|l|}{\textbf{Frame display (ms)}}      & 25.1               & 5.4                 & 27.3               & 4.7                 \\ \hline
\end{tabular}
\end{scriptsize}
\end{table}

A capacitive touch screen is the most common method of controlling a mobile device. Processing the touch events does however incur significant delay to the system. We measured a delay of over 40 ms on the older Samsung S4 while the newer Samsung S7 handles touch commands significantly faster with an average delay of 24 ms. 

User's control commands can also be inputted through an external gamepad or an integrated hardware controller. The USB connection conveys user commands to the operating system considerably faster than the capacitive touch screen. Our measurements show a negligible delay of under 1 ms on both tested devices. The operating system passes on the control command to the running application which receives a callback after a short delay. We measured this delay to be approximately 6 ms on the Samsung Galaxy S4 and roughly 3 ms on the newer S7. The control command is sent to the cloud gaming server after the control has been registered in the cloud gaming application code. Samsung S4 sends the commands in approximately 9 ms while the S7 uses under 2 ms to send a control command to the network.

\subsection{Frame receive and decode}
Android's decoder takes full frame buffers as an input. However the server sends the frames in multiple network packets. The packets have to be buffered in the client side before handing them over to the decoder. We measured a delay of roughly 10 ms for both devices between the first and last packet's arrival of a single frame.

Android has built-in video decoders for h264 video which are used by the GamingAnywhere application. We instrumented the code to measure the time between a frame being inputted to the decoder and the time the frame is decoded and ready to be displayed. With FullHD video (1920x1080) the S4 averaged a delay of 20 ms while the S7 was considerably faster with a delay of 8 ms.

\subsection{Frame display}
Frame display time is the delay before a frame is visible on the screen after it has been handed over from the media decoder to the display buffer. The results show that this delay is not trivial and is actually one of the largest single components affecting the overall delay. The refresh rate of a mobile device's display is usually 60 Hz which translates to roughly 17 ms delay between display updates. Android OS uses double buffering which adds to the overall frame display time.  Depending on how the decoded frames and vsync of the display line up, the frame display time will be between 17-34 ms with an average of 1.5 vsync periods (25 ms). The measurement results confirm this logic in both phones as the display refresh rate is same in both of the tested devices.

\subsection{Network and server delay}
In cloud gaming the game is rendered on a distant server. This adds network and server processing delay on top of the client-side delays. Even in optimal conditions the network delay is at least 20 ms and the server delay around 15 ms. These are naturally highly dependent on the location and processing power of the server infrastructure. We compare the overall latencies discovered further in Chapter \ref{chapter:discussion}.

\subsection{Summary}
Two surprising components dominate the overall client-side delay: touch input processing and frame display. By using an external usb-connected controller, we can reduce the overall latency by over a third with both mobile phones. Frame encoding and decoding is usually associated as the dominating factor in the cloud gaming pipeline. However the decoding part at least seems to be getting faster and faster while the frame display time dominates still with both generations of phones. This could however be mitigated with a method called scanline racing which is introduced in the latest Android operating system (7.0) mainly for virtual reality applications. We measure this feature in Chapter \ref{chapter:virtual_reality}.


%% file: mcglatencypaper-delayanalysis-vr.tex
\label{chapter:virtual_reality}

Virtual reality (VR) applications require a very low delay between user controls and visual feedback. VR applications are usually controlled by head movement and a possible hand-held external controller. The mobile device itself is attached to a head mounted device. In this chapter we measure these VR-application specific control delays and show how the new async reprojection feature of the newest Android operating system could decrease the frame display delay. We use the simple Treasurehunt application from the Android VR SDK samples in our measurements.

\subsection{Control delay}

The industry standard in running virtual reality applications on mobile devices is to attach the device to a wearable headset. The built-in sensors of the mobile phone can be utilized to track the head movements of the user. Additionally a  handheld control device can be used for further user interaction. In the latest VR platform guidelines by Google (Daydream) the control device is a special purpose Bluetooth-device called the Daydream controller. In the measurements we measure the delay of the built-in gyroscope, which is used for head tracking and an Arduino-compatible board with a Bluetooth (BLE) connection to the mobile phone. The Arduino-device mimics the Daydream controller and also provides a reference orientation.  We address the gyro and Bluetooth delays separately as the perceived delay depends on the what the user is performing. The gyro sensor delay is perceived only in head tracking while the bluetooth delay is present in other control commands.

The orientation of the mobile device is received from the gyroscope sensor inside the mobile device. A callback is triggered in the application code when the sensor reading has changed. The callback includes the sensor event and a timestamp of when the event happened. We used the raw values of the Arduino gyro setup explained in Chapter \ref{chapter:measurementsetup} as a reference. The average results of the gyro delay measurements can be seen in Table \ref{table:vrcontroldelay}. On Samsung S7 the timestamps generated for the sensor events were by average 5.8 ms delayed from the reference gyro setup. The callback was fired by average 6.4 ms after the sensor timestamp. Nexus 6P portrayed similar results while the Samsung S4 had significantly more delay in its gyro values.


\begin{table}[]
\centering
\caption{Mobile device gyro sensor and Bluetooth delay.}
\label{table:vrcontroldelay}
\begin{scriptsize}
\begin{tabular}{l|c|c|c|c|}
\cline{2-5}
                                             & \multicolumn{2}{l|}{\textbf{Gyro delay(ms)}} & \multicolumn{2}{l|}{\textbf{Bluetooth delay (ms)}} \\ \hline
\multicolumn{1}{|l|}{\textbf{Mobile device}} & \textbf{Avg.}        & \textbf{SD}       & \textbf{Avg.}           & \textbf{SD}          \\ \hline
\multicolumn{1}{|l|}{Samsung S4}             & 78.8                 & 32.8                  & 17.5                    & 5.2                      \\ \hline
\multicolumn{1}{|l|}{Samsung S7}             & 12.2                 & 4.1                   & 22.0                    & 4.9                      \\ \hline
\multicolumn{1}{|l|}{Nexus 6P}               & 10.2                 & 3.2                   & 28.2                    & 6.5                      \\ \hline
\end{tabular}
\end{scriptsize}
\end{table}

The Bluetooth controller delay was measured from sending an update from the Teensy device to the callback in the Java application code. 
The preferred delay was set to the minimum interval defined in the BLE standard (7.5 ms). Interestingly we observed from the Bluetooth communication logs that the Samsung S4 (Android 5.0) accepted this delay while the Samsung S7 (Android 6.0) and Nexus 6P (Android 7.0) only accepted a slightly larger interval of 11.5 ms. This difference can be observed from the minimum delays recorded. This limitation might be because of energy consumption optimizations in the later Android versions. The measured average delays are presented in Table \ref{table:vrcontroldelay}.

\subsection{Frame draw and display}
In contrast to RDR and MAR applications the entire three-dimensional environment is rendered on the mobile device in virtual reality applications. 
The delay perceived by the user is the difference between the time of a head movement and the time when this information is used in a displayed frame.

We have already established the raw delay in gyroscope values compared to the reference setup. The gyroscope information however still needs to be applied to a rendered view and the frame needs to be displayed through the screen of the mobile device. We injected timing hooks to the sample VR application and timed how long it takes to render a single frame using the setup presented in chapter \ref{chapter:measurementsetup}.

In the latest Android version (7.0) a mode called asynchronous reprojection is available to supported devices. The mode tries to lower the rendering latency by decoupling the framerate from the display framerate. It enables scanline racing where the image is directly rendered to the front buffer just before it is scanned out.

The display latency results measured with the three mobile phones are presented in Table \ref{table:vrrender}. Nexus 6P is at the time of writing the only available device capable of asynchronous reprojection. We measured its delay both with the mode enabled and disabled. The results show that enabling async reprojection significantly lowers the overall display latency. The base line latency without the method enabled is however surprisingly high compared for example to the cloud gaming use case where instead of OpenGL graphics video was rendered to the screen.

\begin{table}[]
\centering
\caption{Frame draw and display delay results for the MVR application.}
\label{table:vrrender}
\begin{scriptsize}
\begin{tabular}{l|c|c|c|c|}
\cline{2-5}
                                             & \multicolumn{2}{c|}{\textbf{Frame draw (ms)}} & \multicolumn{2}{c|}{\textbf{Frame display (ms)}} \\ \hline
\multicolumn{1}{|l|}{\textbf{Mobile device}} & \textbf{Avg.}        & \textbf{SD}        & \textbf{Avg.}          & \textbf{SD}         \\ \hline
\multicolumn{1}{|l|}{Samsung S4}             & 5.2                  & 2.2                    & 55.9                   & 5.0                     \\ \hline
\multicolumn{1}{|l|}{Samsung S7}             & 13.2                 & 2.2                    & 58.3                   & 1.7                     \\ \hline
\multicolumn{1}{|l|}{Nexus 6P}               & 12.3                 & 6.8                    & 67.5                   & 1.3                     \\ \hline
\multicolumn{1}{|l|}{Nexus 6p (async)}       & 12.3                 & 6.8                    & 34.9                   & 7.6                     \\ \hline
\end{tabular}
\end{scriptsize}
\end{table}

%% file: mcglatencypaper-delayanalysis-ar.tex
Mobile Augmented Reality (MAR) is another application type requiring a low-latency pipeline for an acceptable QoE. 
 AR applications differ from the RGR and the MVR scenario by using the camera as the main input as the augmented content is added on top of a projection of the real world. We measured the delays with the ARToolkit sample app ARSimpleProj which draws a cube on top of a marker.


\subsection{Control delay}

Initial delay present in any MAR application is the delay between the image sensor starting to capture an image and the capture result callback in the application code. We measured this delay with the Samsung S7 and Nexus 6P mobile devices using the Camera2 API introduced in Android 5.0. Samsung S4 uses an unspecified starting point in its camera timestamps and was left out from the measurements. We used resolutions 1440x1080 and 640x480 with 30 fps frame rate in all tests as they were the highest and lowest resolutions supported by both phones. The results are presented in Table \ref{table:arrender}. Both tested devices were similar in performance with a delay of 60 to 90 ms depending on the resolution.

\begin{table}[]
\centering
\caption{Camera frame, frame draw and frame display delay results for the MAR application.}
\label{table:arrender}
\begin{scriptsize}
\begin{tabular}{l|c|c|c|c|c|c|}
\cline{2-7}
                                             & \multicolumn{2}{c|}{\textbf{\begin{tabular}[c]{@{}c@{}}Camera delay\\ (640p/1440p) (ms)\end{tabular}}} & \multicolumn{2}{c|}{\textbf{\begin{tabular}[c]{@{}c@{}}Frame\\ draw (ms)\end{tabular}}} & \multicolumn{2}{c|}{\textbf{\begin{tabular}[c]{@{}c@{}}Frame\\ display (ms)\end{tabular}}} \\ \hline
\multicolumn{1}{|l|}{\textbf{Mobile device}} & \textbf{Avg.}                                       & \textbf{SD}                                      & \textbf{Avg.}                               & \textbf{SD}                               & \textbf{Avg.}                                 & \textbf{SD}                                \\ \hline
\multicolumn{1}{|l|}{Samsung S7}             & 63.5/88.6                                           & 1.9/3.9                                          & 16.8                                        & 6.2                                       & 36.7                                          & 5.0                                        \\ \hline
\multicolumn{1}{|l|}{Nexus 6P}               & 60.0/78.8                                           & 1.9/3.3                                          & 20.0                                        & 6.1                                       & 36.9                                          & 5.1                                        \\ \hline
\end{tabular}
\end{scriptsize}
\end{table}

\subsection{Frame draw and display}
A typical MAR application combines the input frame from the camera with a rendered object. The location of the object marker needs to be tracked for each frame. Table \ref{table:arrender} shows the measured frame draw and display times. Frame draw times include search for the location of the marker in the camera frame and all the draw commands in the application code. Frame display time is the delay after the frame is drawn in the code and the time the frame is displayed on the screen of the mobile device.

\subsection{Offloading example}

The overall delay of a MAR application is highly dependant on the amount of processing needed to display the result. Some tasks have to be offloaded to an external server. For this use case we analyzed an application which plays a movie trailer when the mobile phone's camera is pointed towards a movie poster. The trailer is overlaid according to the homography of the posters. The poster recognition and pose estimation is done on the server and the mobile client keeps tracking and estimating the pose of the poster after receiving the result. For availability reasons we used the Xiaomi Mi 5 mobile phone for measuring the MAR offloading application delay. The phone's performance should be similar to the Samsung S7.

The application offloads image recognition to an external server. This means every 60th frame is sent out and the result is used to track the object on the mobile device. This delay from the application to the server and back was measured to be approximately 500 ms. The constant delay visible to the user is however the local tracking time of the object. We measured this delay to be 24 ms on average which is line with the sample application measurements.

%% file: mcglatencypaper-discussion.tex
\label{chapter:discussion}

\begin{figure}[t]
\centering
\includegraphics[width=\columnwidth,height=0.6\columnwidth]{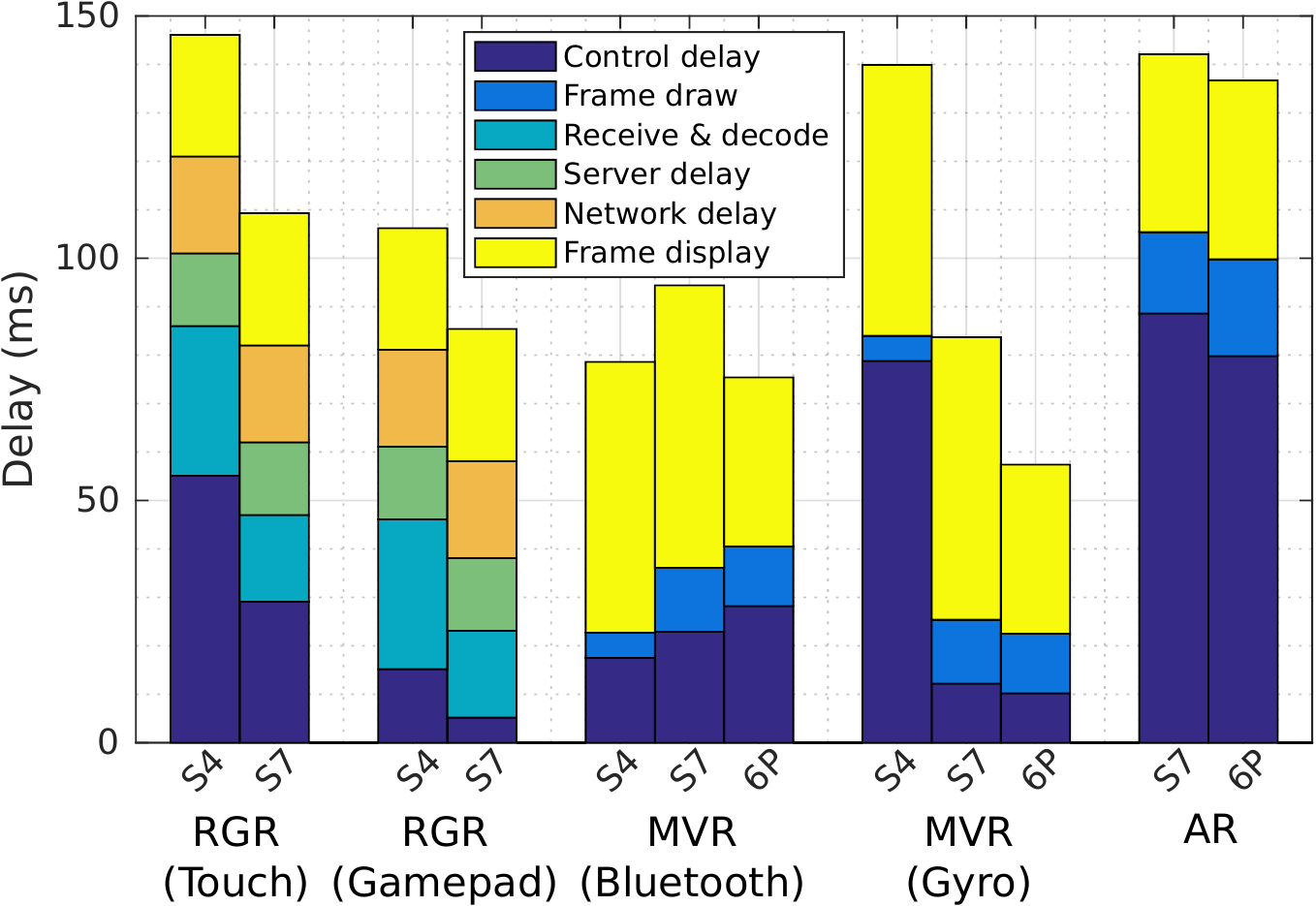}
\caption{Summary of the measured delay scenarios.} 
\label{fig:all_delays}
\end{figure}

The measurements show that the two major components affecting the overall delay in latency-critical mobile video applications are control delay and frame display. This can be observed in the summary presented in Figure \ref{fig:all_delays}. The magnitude of the control delay is highly dependent on the type of input used by the application. A modern mobile device can send gamepad commands to a remote server in a matter of milliseconds while an AR application can wait up to 90 ms to even get a frame from the camera for processing. Top-of-the-line mobile phone (Samsung S7) can process touch and Bluetooth events in roughly 20 to 30 ms while the gyro sensor events arrive faster with an average of 12 ms of delay. Touch screen delays seem to get lower with each mobile phone generation. The gyro sensor delays is also very small with recent mobile phones. The camera feed to the application would however benefit from further optimizations perhaps with the cost of image quality.

Drawing and displaying a single frame is processed in approximately 25 ms in the RGR scenario where a video stream is received and decoded. Rendering a scene takes longer, we measured a delay of almost 60 ms to display a VR scene for both eyes in a head-mounted setup. Using the latest features of the Android OS, this can be lowered to 35 ms.

While deep understanding of the impact of latency on user experience is still an open problem, previous research has discovered many things about human perception of latency with modern mobile technology. For example, Deber et al. recently characterized the Just Noticeable Difference (JND) and the impact of additional latency on task performance in direct and indirect user interaction with a touch device\cite{deber2015much}. They found that the mean JND for a simple tapping task is 69 ms and 96 ms for direct and indirect touch, respectively. The JND is substantially shorter when performing a dragging task. In our target applications, mobile cloud gaming in particular, pressing virtual buttons on a touch screen can be argued to be a combination of the two. A comparison of these JND numbers to the results on touch, gamepad, and Bluetooth-based controls in Figure \ref{fig:all_delays} reveals that only the VR case with Bluetooth-based control can reach end-to-end latency below the JND limit of tapping latency in the best case. Notice that mobile VR with offloaded graphics rendering, in case it is too computationally demanding for the mobile device to perform, is essentially the RGR case, only the control delay being different.

Lee et al. studied error rates in pointing tasks where a target is about to appear within a limited time window for selection\cite{lee2016chi}.
They found out that variability in the timing of the pointing event with a touch screen causes lower user performance in gaming compared to using physical keys, for example. 
One interesting observation from our results related to the variability is the effect of asynchronous reprojection: It lowers the latency on average but it also increases its variance. However, further user studies are required to quantify its exact effect on user experience. We also note that using an external gamepad instead of a touch screen clearly reduces both the average latency and its variance.   

Also application-specific studies have been performed. \\Among the RGR applications, cloud gaming has been investigated. 
Subjective tests show a clear correlation between QoE and latency but it is difficult to precisely quantify their relationship\cite{jarschel2011evaluation,lee2012all,claypool2014effects,raaen2014can,clinch2012percom}. Unfortunately, most of the studies have not measured the true end-to-end latency because of which the results are difficult to interpret and compare to the latencies we have measured.

Acceptable delay for VR applications is also a debatable question. Previous research \cite{jerald2010scene} has shown that the threshold for latency is very subjective, some hardly notice a delay of 100 ms while others can perceive delays down to 3-4 ms. Also velocity of head movements influence the tolerance of delay\cite{allison2001vr}. Some industry representatives have stated that a latency lower than 20 ms is recommended \cite{carmack_blog,abrash}. Our measurements show that the recent features in the Android operating system enable the overall latency to go under 50 ms for a compatible device. This shows that there is still room for improvement in mobile device hardware to achieve seamless interaction with the user. The measurements show that more powerful GPUs and displays with higher refresh rates could alleviate the end-to-end delay substantially. Another option is to offload the graphics rendering to an external server. The network delay with current technologies might however increase the latency even further.

The quest for low latency has mostly focused earlier on making the network delay shorter through novel architectures. For example, Satyanarayanan et al. have presented Cloudlets that offer computing power for mobile clients within one-hop latency\cite{satyanarayanan2009case}. A more incremental approach by Choy et al. is to use the existing CDN network to offload computation from the mobile device\cite{choy2014hybrid}. However, as Figure \ref{fig:all_delays} shows, network latency together with access delay is only part of the delay pipeline in latency-critical mobile video applications. Processing delay on the server side as well as on the client side occur both in the software and hardware. Jain et al. focus on balancing the network and the computational delay with accuracy in mobile AR\cite{jain2015mobisys}.
Lee et al. took a different path and developed a system to speculatively execute different possible scenarios of a cloud game in order to mask latency\cite{lee2015mobisys}. In a similar vein, Boos et al. built a system to aggressively precompute and cache all possible images that a VR user might encounter in order to achieve low latency and energy consumption\cite{boos2016mobisys}. These solutions are useful for minimizing the latency components excluding the control and the frame draw \& display ones. 

Finally, we point out that multiuser scenarios introduce additional latency components due to geographic distance between users and our current study excludes those.

%% file: mcglatencypaper-conclusion.tex
We presented a measurement methodology to study the latency within a mobile device and apply it to three different interactive mobile multimedia applications for which low latency is very important for the user experience. Our results demonstrate that the delays vary substantially between device models, applications, and input methods used. 
Comparing our results to those obtained with user studies on the effect of latency, the technology does not appear to be mature enough yet for completely seamless user experience.